\documentclass[doublecol]{epl2} 

\title{Electron pockets
and  pseudogap Dirac point in underdoped
cuprate superconductors}
\shorttitle{Electron pockets and pseudogap Dirac point} 

\author{K.A.G. Fisher\inst{1,2} \and E.J. Nicol\inst{1,2} \and J.P. Carbotte\inst{3,4}}
\shortauthor{K.A.G. Fisher \etal}
\institute{
\inst{1}Department of Physics, University of Guelph,
Guelph, Ontario, Canada N1G 2W1\\
\inst{2}Guelph-Waterloo Physics Institute,
University of Guelph, Guelph, Ontario, Canada N1G 2W1\\
\inst{3}Department of Physics and Astronomy, McMaster
University, Hamilton, Ontario, Canada L8S 4M1\\
\inst{4}The Canadian
Institute for Advanced Research, Toronto, Ontario, Canada M5G 1Z8
}
\pacs{74.72.-h}{Cuprate superconductors}
\pacs{74.72.Kf}{Pseudogap regime}
\pacs{74.20.Mn}{Nonconventional mechanisms (spin fluctuations, polarons and bipolarons, resonating valence bond model, anyon mechanism, marginal Fermi liquid, Luttinger liquid, etc.)}

\abstract{
We consider a model of the pseudogap specifically designed to describe
the underdoped cuprates and which exhibits particle-hole asymmetry.
The presence of electron pockets, besides the usual hole
pockets, leads to the appearance of new vectors beyond the
usual so-called octet model in the joint density of states (JDOS), 
which underlies the analysis of Fourier-transform scanning tunneling
spectroscopy (FT-STS) data. These new vectors are associated with
distinct patterns of large amplitude in the JDOS 
and are expected to occur primarily at positive bias. Likewise a pseudogap
Dirac point occurs at positive bias and
this point can be determined either through FT-STS
or through extrapolation of data from
the autocorrelation function of angle-resolved photoemission spectroscopy.}

\begin{document}

\maketitle

\begin{figure}
\includegraphics[width=0.48\textwidth]{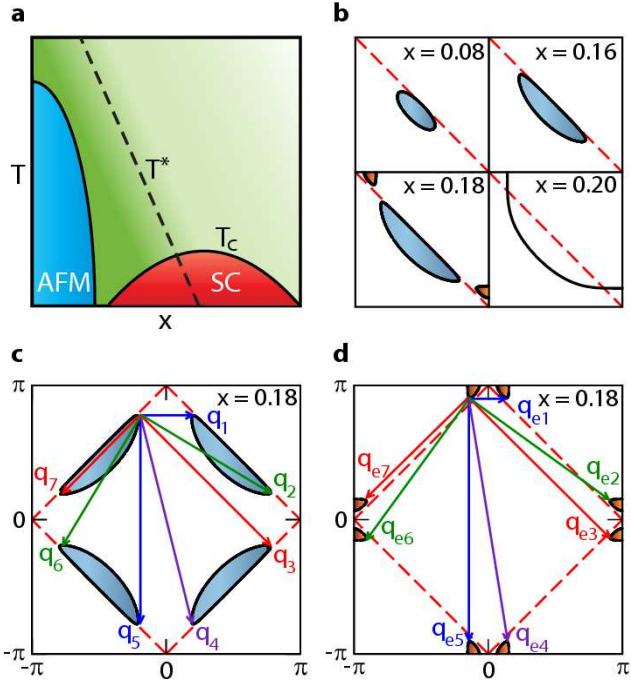}
\caption{(Colour online)
(a) Generic phase diagram of the hole-doped cuprates
showing the antiferromagnetic Mott insulator (AFM) region, the superconducting
dome (SC) and the $T^*$ line indicating the onset of the pseudogap phase
at lower doping $x$ and temperature $T$. 
(b) The reconstruction of the Fermi surface in the first quadrant of the
square Brillouin zone as the doping is varied from
underdoped ($x=0.08$, 0.16, 0.18) to optimally doped ($x=0.2$). 
When no pseudogap
is present, as in the case of $x=0.2$ which is taken as the quantum critical point where
the $T^*$ line goes to zero, then there is  a large Fermi surface. With slight
underdoping electron pockets (red) and a hole pocket (blue) appear about the
AFM Brillouin zone boundary (red dashed line). These pockets shrink with underdoping.
(c) The octet model  $\boldsymbol{q}$'s  of AC-ARPES
and FT-STS which connect points of high density of states associated with
the hole pockets (or arcs in some models).
(d) The electron pockets for $x=0.18$ 
will also be connected by eight $\boldsymbol{q_e}$'s.
Not shown here are possible interband $\boldsymbol{q}$'s which are in addition
to the intraband $\boldsymbol{q}$'s shown in (c) and (d).
}
\label{fig1}
\end{figure}

Prominent in the
underdoped cuprate superconductors
 is the existence of a pseudogap 
in the excitation spectrum which opens above $T_c$
but below a temperature $T^*$.\cite{statt} 
Whether this gap is the same as the superconducting energy gap
or is a manifestation
of a competing order independent of the superconductivity remains
an open and central question. Indeed,
it is not clear that the two energy scales seen in experiments\cite{sacuto,hudson,huefner} 
arise from two separate gaps\cite{norman}. 
If there are two distinct gaps
of $d$-wave symmetry,
they each will exhibit a Dirac point, a point at which 
linear dispersions cross, 
at a different energy and momentum in the band structure.
Evidence for a single gap would favour theoretical proposals of preformed Cooper
pairs above $T_c$  which phase lock at this
temperature.\cite{emery} On the other hand, models with
competing order typically have a second energy gap scale and some form
of Fermi surface reconstruction which gives rise to the possibility of
hole and electron pockets.\cite{laughlin,zhu} 
While hole pockets or arcs have been seen in ARPES
experiments,\cite{arcs,holes,hbyang2010} 
recently the observation of electron pockets from quantum oscillations
in high magnetic
fields has been reported.\cite{louis2}
 Furthermore, the preformed pair theory gives
rise to particle-hole symmetric quantities whereas competing orders can
exhibit particle-hole asymmetry.\cite{hashimoto,adrien}
 As two-gap scenarios typically arise from
strong correlations due to the nearby presence of the antiferromagnetic
Mott insulating state shown in the phase diagram of fig.~\ref{fig1}(a), the
AFM Brillouin zone boundary plays a significant role in determining the
reformation of the large Fermi surface to small pockets with underdoping as
shown in fig.~\ref{fig1}(b). Given that
 the underlying mechanism of superconductivity is most
likely to be attributed to antiferromagnetic spin fluctuations, as directly
evidenced
by experiment\cite{jules}, both the pseudogap and superconducting gap are expected to reflect
the same symmetry which is known to be $d$-wave in the cuprates\cite{kirtley}. Thus, the
gaps will exhibit Dirac points in the $(\pi,\pi)$ direction with that
for superconductivity on the Fermi surface (taken as $\omega=0$ in energy)
and that for the pseudogap shifted to positive energy and at a different
$\boldsymbol{k}$ point in the Brillouin zone\cite{adrien}. 
In this letter, we
demonstrate that the existence of electron pockets
will lead to additional characteristic features in the
quasiparticle interference (QPI) patterns\cite{qpi1,qpi2} 
obtained from Fourier transform
scanning tunneling spectroscopy (FT-STS).
Furthermore, we discuss the various signatures of a particle-hole asymmetric
pseudogap that may lead to the possible identification of the pseudogap
Dirac point through FT-STS or via
the extrapolation
of information
from the autocorrelation
function of angle-resolved photoemission spectroscopy (AC-ARPES)\cite{mcelroy,acarpes}.

It was shown\cite{mcelroy} that there is a direct correspondence between the
QPI measured in FT-STS and the AC-ARPES. This important work confirmed that
an evaluation of the joint density of states (JDOS) in terms
of the spectral function $A(\boldsymbol{k},\omega)$:
\begin{equation}
JDOS(\boldsymbol{q},\omega)=\frac{1}{V}\sum_{{\boldsymbol{k}}}A({\boldsymbol{k}},\omega)
A({\boldsymbol{k+q}},\omega)
\end{equation}
will give rise at fixed energy $\omega$ to a series of strong peaks at the tips of eight
${\boldsymbol{q}}$ vectors (one of which is 0) which is known as the octet
model\cite{octet}. For a quasiparticle dispersion given by $E_{\boldsymbol{k}}$,
these $\boldsymbol{q}$'s map a point of high density of states due to
$1/\nabla E_{\boldsymbol{k}}$ to another such point as shown in fig.~\ref{fig1}(c). The patterns 
for the JDOS formed experimentally by an autocorrelation function of the
ARPES $A(\boldsymbol{k},\omega)$ agree well with 
the QPI found in FT-STS, which is based on
the same physics.\cite{mcelroy} 
We now show from the JDOS that 
 AC-ARPES, indirectly, and the QPI FT-STS, more directly, 
can provide a unique way for determining
both the energy and momentum information associated with the pseudogap
Dirac point and also electron pockets.

\begin{figure}
\includegraphics[width=0.48\textwidth]{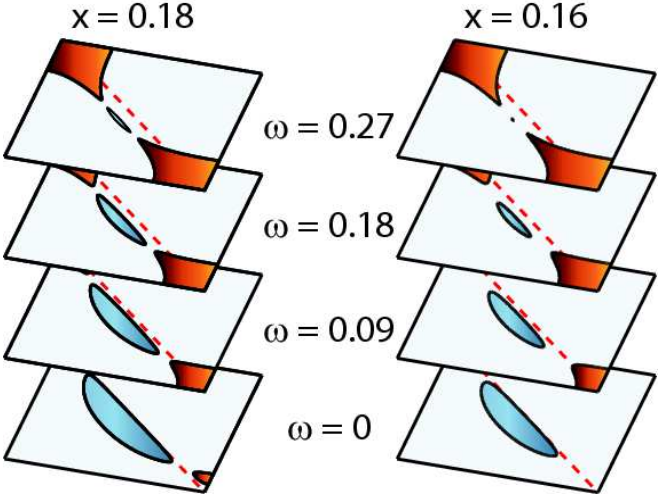}
\caption{(Colour online) Illustration of the evolution of 
hole and electron pockets for
positive energy cuts of the energy bands. For $x=0.18$, electron pockets
exist at $\omega=0$ and grow as the hole pocket shrinks. The $x=0.16$ case
demonstrates that even if electron pockets do not exist at $\omega=0$,
they will appear at higher positive 
bias. For $x=0.16$, at the largest energy shown,
{\it i.e.} $\omega=0.27$ in units of $t_0$, the hole pocket has shrunk to
a point which is the pseudogap Dirac point. For $x=0.18$, this will be
reached at greater energy.
}
\label{fig2}
\end{figure}

To provide a concrete demonstration of the expected signatures of a second Dirac
point, electron pockets and particle-hole asymmetry, we adopt the model
of Yang, Rice and Zhang\cite{yrz} who provide a phenomenological ansatz for a
Green's function which represents a pseudogap state in the cuprates.
This model has been developed out of prior work on t-J models, Hubbard ladders
and resonating valence bond spin liquid 
theory\cite{yrz} and has proven to be quite effective at describing
a large body of experimental 
data\cite{yrz,adrien,adriencv,emilia,kent,yrzarpes,jamescv,bascones,yrzandreev,james2011,adam,belen,jamesraman,hbyang2010,khodas,yrzcheckerboard}.
In this model, the coherent part of the Green's function in the pseudogap
state has the form of 
\begin{equation}
G({\boldsymbol{k}},\omega)
=\frac{g_t}{\omega-\epsilon({\boldsymbol{k}})-\Sigma_{\rm pg}({\boldsymbol{k}},\omega)} ,
\label{eq:G}
\end{equation}
where the electron self-energy is given by
$\Sigma_{\rm pg}({\boldsymbol{k}},\omega)=|\Delta_{\rm pg}({\boldsymbol{k}})|^2/[\omega+\epsilon_0({\boldsymbol{k}})]$.
Here, $g_t=2x/(1+x)$
is a Gutzwiller factor that reflects strong correlations and a reduction in
coherence with the approach toward the AFM Mott insulator. The pseudogap
is taken to be $\Delta_{\rm pg}({\boldsymbol{k}})=\Delta_{\rm pg}(x)[\cos(k_xa)-\cos(k_ya)]/2$,
with $\Delta_{\rm pg}(x)=0.6(1-x/0.2)$. The band structure $\epsilon({\boldsymbol{k}})$
is taken to be a third nearest-neighbour tight-binding dispersion and
$\epsilon_0({\boldsymbol{k}})=-2t_0[\cos(k_xa)+\cos(k_ya)]$, where $\epsilon_0({\boldsymbol{k}})=0$
gives the AFM Brillouin zone boundary shown in fig.~\ref{fig1}. Indeed, it is the fact
that $\epsilon_0({\boldsymbol{k}})$ appears
in the electron self-energy $\Sigma_{\rm pg}({\boldsymbol{k}},\omega)$, 
instead of $\epsilon({\boldsymbol{k}})$, that
the pseudogap opens away from the Fermi surface $\epsilon({\boldsymbol{k}})=0$
on a surface defined by $\epsilon({\boldsymbol{k}})+\epsilon_0({\boldsymbol{k}})=0$,
giving rise to
particle-hole asymmetry in this model. Further modifications
of this form
for a superconducting gap $\Delta_{\rm sc}({\boldsymbol{k}})=\Delta_{\rm sc}(x)[\cos(k_xa)-\cos(k_ya)]/2$, where $\Delta_{\rm sc}(x)=0.14[1-82.6(x-0.2)^2]$, are standard and given
in refs.~\cite{yrzarpes,adrien}, 
where the typical band structure parameters used here may also be found.
{(Note, that prior work on the $JDOS$ has been done in this 
model\cite{bascones}, however, a theory was used 
for the superconducting state which has
since been revised\cite{yrzarpes} and so results of ref.~\cite{bascones}
differ from ours
in the presence of superconductivity.)}
While we use a particular model for illustration, other models with
particle-hole asymmetry would also give qualitatively similar results. 
The pockets of fig.~\ref{fig1}(b) are a result of this
model and it should be noted that at the Fermi energy the AFM Brillouin zone
side has small quasiparticle
weight and hence these pockets can appear as arcs in 
experiment. Experimental evidence for these types of pockets in agreement
with this theory has been recently presented in ref.~\cite{hbyang2010}. 
Figure~\ref{fig1}(d) illustrates that should electron pockets exist, they will also
give rise to new peaks in the JDOS which would be seen in experiment. Note
that while it might appear that only a narrow range of doping gives rise to electron pockets,
as seen in fig.~\ref{fig1}(b),
it is shown in fig.~\ref{fig2} that electron pockets will appear at finite bias on
the positive side even when not present at $\omega=0$.
Moreover, due to the formation of Bogoliubov quasiparticles in the
superconducting state, an image of these pockets can appear at
negative bias.\cite{adrien,james2011}
The pseudogap Dirac point is seen in this figure as the single point in the
top right hand sheet for $x=0.16$. It will be found at positive energy
along the nodal direction about halfway in momentum
between the Fermi surface  at $\omega=0$ and the AFM
Brillouin zone.\cite{adrien} The observation of electron pockets and the pseudogap
Dirac point at positive bias poses a problem for ARPES which sees only the 
occupied states at negative energies, however, ARPES can potentially be used
to obtain very useful but indirect information on
 the Dirac point from the particle-hole asymmetry.

\begin{figure}
\includegraphics[width=0.48\textwidth]{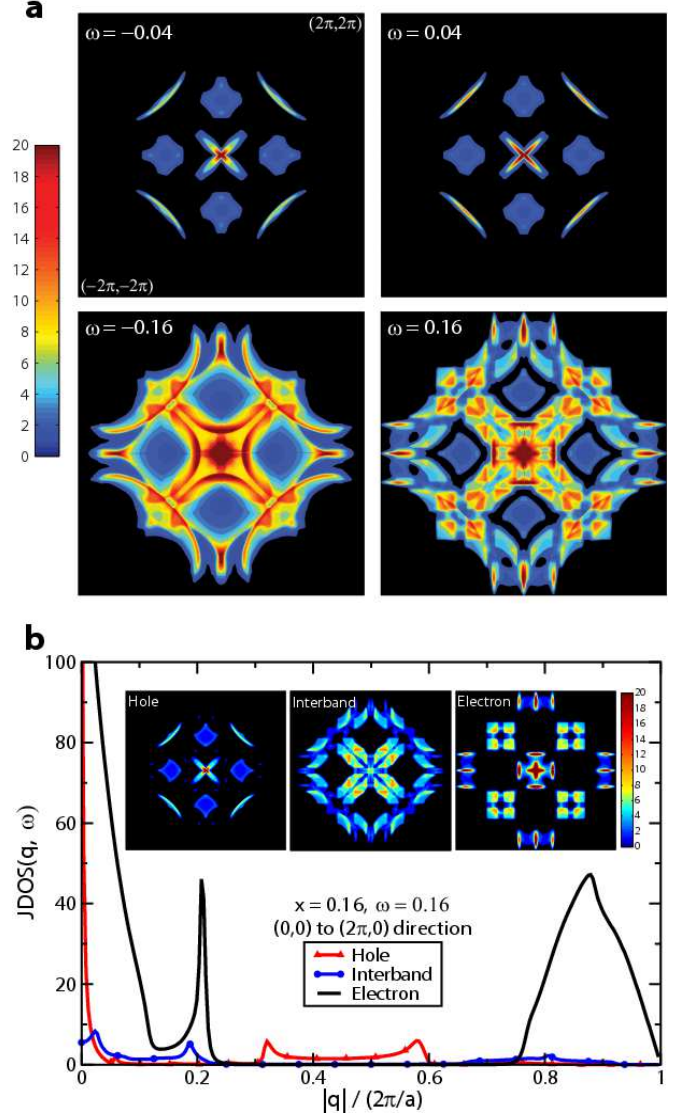}
\caption{(Colour online)
 (a)
The JDOS in the superconducting state with the pseudogap present
for $x=0.16$. Each colour map is shown for $\boldsymbol{q}$ varying from
$(-2\pi,-2\pi)$ in the lower left corner to $(2\pi,2\pi)$ in the upper
right corner of the square. The left (right) frames are for negative (positive)
bias. The top frames are for $|\omega|=0.04t_0$ which is below the superconducting
gap maximum $\Delta_{\rm sc}=0.12t_0$. 
The bottom two frames are for
$|\omega|=0.16t_0$, well above $\Delta_{\rm sc}$.
(b) The JDOS in the (0,0) to ($2\pi$,0) direction showing intraband (hole
and electron) and interband contributions to the autocorrelation function
at energy $\omega=0.16t_0$.
The inset shows the JDOS
in all directions of $\boldsymbol{q}$ decomposed into the three components.
}
\label{fig3}
\end{figure}

Clearly, with the Dirac point at positive energy, an asymmetry
exists in this situation. To discuss this point, we show in fig.~\ref{fig3} the
particle-hole asymmetry that exists in the JDOS. While our prior figures were
solely about the pseudogap state, this figure now includes superconductivity.
In the pure pseudogap state, particle-hole asymmetry would occur at all biases but
the presence of superconductivity restores the symmetry for energies below
the relevant superconducting energy scale. 
This is seen in fig.~\ref{fig3}(a) where the top frames show the JDOS at
positive and negative biases and the energy $|\omega|<\Delta_{\rm sc}$. 
These two frames show the particle-hole
symmetry imposed by superconductivity. The bottom two frames for  
$|\omega|>\Delta_{\rm sc}$ illustrate the particle-hole asymmetry due to
the pseudogap state. Using the case of $\omega=0.16t_0$, fig.~\ref{fig3}(b)
shows a decomposition of the colour map into its contributions for the hole
intraband,
electron-hole interband, and electron intraband pieces, respectively (see inset)and the strength of the JDOS for a cut along the center for $\boldsymbol{q}=(0,0)$ to $(2\pi,0)$.
For this direction, the electron JDOS contribution dominates and the
interband component is not important. Note that the JDOS shown here should be scaled by
a factor of $g_t^2/4\pi^2$ in the model used here.
The inset colour maps confirm the hole JDOS to be as 
shown before in prior works but in addition illustrate
a feathered pattern for the interband JDOS and a distinctive 4-square pattern
for the electron JDOS which also dominates the full $\boldsymbol{q}$-map in the
lower right frame of (a). Indeed, one sees that there are very different
characteristic fingerprints of electron pockets versus hole or interband
and while much attention has been paid to the hole pockets, these other
features can be distinguished in the composite colour map
for $\omega=0.16$ in fig.~\ref{fig3}(a).
Here, as we will focus on the JDOS for
$\boldsymbol{q}$ varying from $(0,0)$ to $(2\pi,0)$ only for brevity,
we emphasize that the potential interband JDOS peaks are small in this
direction and our discussion can focus solely on the independent
electron and hole features. 

\begin{figure}
\includegraphics[width=0.48\textwidth]{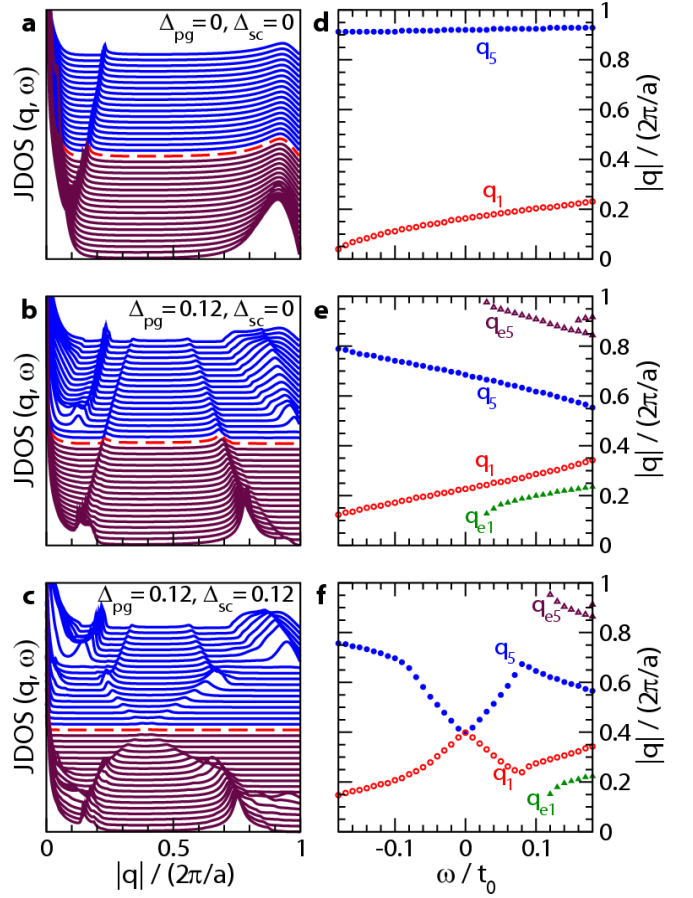}
\caption{(Colour online) Signatures of electron pockets in the 
autocorrelation function.
The JDOS for $x=0.16$ in the $\boldsymbol{q}=(0,0)$ to ($2\pi,0$) direction
for the case of: (a) no gaps (a simple Fermi liquid), (b) 
$\Delta_{\rm pg}$
finite and $\Delta_{\rm sc}=0$ and (c) in the superconducting state with
both gaps finite. Curves are shown offset from one another for negative bias
(purple) to positive bias (blue) with steps of $\omega=0.01t_0$. The dashed curve
indicates the $\omega=0$ case. The variation of $\boldsymbol{q}$ versus
$\omega$ following the main peaks in (a)-(c)  are shown in (d)-(f), 
respectively. 
New $\boldsymbol{q}$'s arise at positive bias indicating
the existence of the electron pockets, and the Dirac point for the superconducting
gap is seen in (f) at $\omega=0$. See text for further discussion.
}
\label{fig4}
\end{figure}

Figure~\ref{fig4} emphasizes the typical JDOS curves and ${\bf q}$ dispersions
for the case of no gaps, with a pseudogap and for both gaps present. Particle-hole
asymmetry is imposed by the pseudogap and electron pockets give rise to 
new $\boldsymbol{q}$'s at positive bias. {The pseudogap-only case
was obtained previously in ref.~\cite{bascones} for the negative bias
range of $\omega=-0.1t_0$ to 0 and our results appear to be in agreement
in that range. These authors refer to this behaviour as non-dispersive 
but we find it to be dispersive on our expanded scale. }
With superconductivity, the 
$\boldsymbol{q_5}$ and
$\boldsymbol{q_1}$ form an X-structure in the $\boldsymbol{q}$ versus $\omega$
plot with crossing point at $\omega=0$. This is the superconducting Dirac point
and it occurs at a specific value of $|\boldsymbol{q}|$ which identifies
the Fermi momentum $k_F=|\boldsymbol{q}|/\sqrt{2}$ in the nodal direction.
The energy onset of the X is set by the maximum of the superconducting
gap on the hole pocket while the onset of the electron 
$\boldsymbol{q_{e1}}$ and
$\boldsymbol{q_{e5}}$ approximately corresponds to the pseudogap energy 
and is more rigorously identified as an energy scale $\Delta^+_{\rm pg}$ 
discussed in ref.~\cite{adrien}. 
Note that in fig.~\ref{fig4}(e), the $\boldsymbol{q_1}$ and
$\boldsymbol{q_5}$ merge towards a point at positive bias and in spite of the distortion
due to superconductivity in fig.~\ref{fig4}(f), the same overall trend in 
$\boldsymbol{q_1}$ and
$\boldsymbol{q_5}$ 
is maintained. This point of merging is the pseudogap Dirac point.
As a final comment, 
the superconducting and pseudogap Dirac point occur at mid range values
of $\boldsymbol{q}$ concurrent with the nodal direction,
whereas
the electron pocket $\boldsymbol{q}$'s are either large or small in magnitude 
as expected for features in the JDOS which arise from the near antinodal direction.

\begin{figure}
\includegraphics[width=0.48\textwidth]{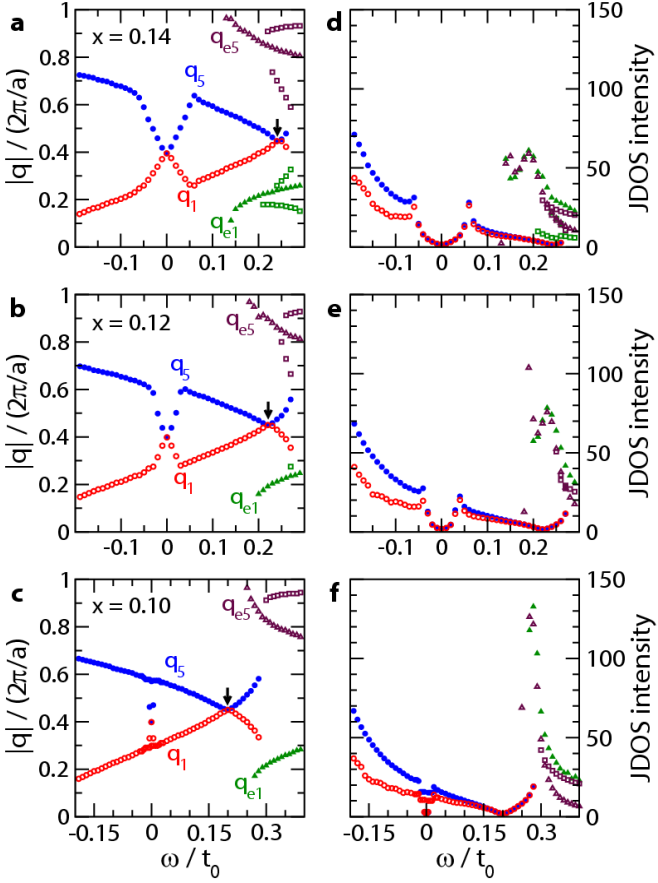}
\caption{(Colour online) 
Evolution with doping of the pseudogap Dirac point.
(a)-(c) Various dopings for the underdoped phase showing  
$\boldsymbol{q}$ versus $\omega$ for the (0,0) to ($2\pi$,0) direction.
In these figures, both the superconducting and pseudogap Dirac points
are visible with the one for the pseudogap (marked by arrows)
 found at positive bias and slightly
different $\boldsymbol{q}$. Extrapolation of the $\boldsymbol{q_1}$
and $\boldsymbol{q_5}$ dispersions from negative bias (ignoring the
X feature due to superconductivity) until the two intercept, should
identify the pseudogap Dirac point. In the right hand frames, (d)-(f),
 the JDOS intensity is plotted for the $\boldsymbol{q}$'s shown
in {\bf a-c}, respectively. The electron pockets provide a strong
signal.
}
\label{fig5}
\end{figure}

Figure~\ref{fig5} emphasizes the pseudogap Dirac point more clearly where for stronger
underdoping the Dirac point shifts to smaller energy as tracked by the black arrows in frames (a)-(c).
Likewise, this point is at a different $|\boldsymbol{q}|$ than the superconducting Dirac point. With nearest neighbour hopping $t_0$ taken to be 150-350 meV
in fitting this model to other experiments,\cite{adrien,kent,yrzandreev,james2011} the pseudogap Dirac point could
shift to energies on order of 
30-70 meV for underdoped Bi-based cuprates with $T_c$
of 45K or less. However, as shown in frames (d)-(f), the intensity at the
two Dirac points falls to zero as the density of states is vanishing at these
points and so the center of the X-feature will fall below
 experimental resolution. Concurrent with this may be matrix element issues
which have been shown to affect the proper observation of the superconducting
Dirac point.\cite{hashimotoprl}
Nonetheless,
one can extrapolate the trend  of 
$\boldsymbol{q_1}$ and
$\boldsymbol{q_5}$ 
to find the point of intersection. While STS experiments can access both
negative and positive bias to form the 
$\boldsymbol{q}$  versus $\omega$ dispersions, currently the positive
and negative bias conductance
is ratioed as $g({\boldsymbol{r}},+E)/g({\boldsymbol{r}},-E)$
 before the Fourier transform is taken.\cite{hanaguri,kohsaka}
 This automatically gives the particle-hole symmetric
${\boldsymbol{q}}$'s relevant for energies below the
superconducting gap. To see the particle-hole asymmetry discussed here,
the positive and negative biases must be kept separate. 
{In older experimental
work showing both positive and negative bias
separately in a few cases,
the data 
is given for
an energy below the superconducting gap energy where particle-hole symmetry
is restored\cite{qpi2,vershinin}.  In ref.~\cite{mcelroy2005}, 
using an analysis to
isolate pseudogap-only regions from an inhomogeneous gap map, 
a non-dispersive $\boldsymbol{q^*}$ is identified
above the superconducting gap energy which is associated with a local
``checkerboard'' charge ordering. No other 
$\boldsymbol{q}$'s  in the pseudogap state are found although the
$\boldsymbol{q^*}$ is not without some dispersion and is close to the
$\boldsymbol{q_1}$ value as illustrated in ref.~\cite{campuzano}.
More recently, this $\boldsymbol{q^*}$ feature has been measured at postive
and negative bias in FT-STS 
for different dopings and temperatures
[fig.~S2(a)-(c) of ref.~\cite{parker2010}] and has been found to disperse
in the pseudogap state. This dispersion
 is in close agreement with the $\boldsymbol{q_1}$
shown here for a reasonable choice of $t_0$.}   
On the other hand,
ARPES can only measure the negative
bias. However, due to the particle-hole asymmetry, the ARPES data for 
$\boldsymbol{q}$  versus $\omega$ can be extrapolated to positive bias
to give an estimate of  the pseudogap Dirac point provided the extrapolation is
taken from above the superconducting energy gap scale.
 Using the published AC-ARPES data presented in
ref.~\cite{mcelroy} to do such an extrapolation, we note
that in the one
underdoped sample shown there are  
kinks in $\boldsymbol{q_5}$, $\boldsymbol{q_7}$ and $\boldsymbol{q_3}$
around 22-25 meV, an indication of a 
possible superconducting gap energy scale.
Extrapolating from energies just above this scale, we roughly estimate that 
$\boldsymbol{q_1}$ and 
$\boldsymbol{q_5}$ intersect at about $\omega\sim 
+40-55$ meV and $|\boldsymbol{q}|/(2\pi/a)\sim 0.45-0.5$ 
(consistent with the results shown here). Alternatively, using $t_0=350$ meV in our calculations to approximately match the kink energies, we infer the Dirac
point to be at $\sim 85$ meV, indicating that the extrapolation from
negative energy is likely to underestimate the energy of the pseudogap 
Dirac point. With either method, this data
might be interpreted to support the presence
of a pseudogap Dirac point and provides evidence for a two gap scenario.
{We note, however, that other AC-ARPES data on 
samples of higher doping and measured at higher temperature find a lack of
dispersion in the pseudogap state\cite{acarpes,campuzano},
which they attribute to $\boldsymbol{q}$'s connecting the tips
of the energy-independent Fermi arcs, rather than charge ordering. 
These
authors also suggest that anomalous behaviour may be seen in the
dispersion of $\boldsymbol{q}$'s in the superconducting state due to electrons coupling to 
collective excitations.}
Clearly additional study by both ARPES and STS communities could further
clarify these important issues. One thing that is clear from fig.~\ref{fig5} is that
electron pockets should provide a strong signal.

At this point, we should address further the assumption 
of particle-hole asymmetry
which is key to our discussion. In addition to our analysis above,
there is recent experimental evidence for particle-hole asymmetry in the
pseudogap state from
both ARPES\cite{hashimoto,hbyang,yrzarpes} 
and STM experiments\cite{pushp,adrien,parker2010}. 
As we have discussed, superconductivity 
largely restores particle-hole symmetry at energies below the superconducting
energy gap scale and hence the potential for detection of 
particle-hole asymmetry in experiment
in the past may have been partly clouded by this issue. 
More recently, however, the evidence has become clearer\cite{hashimoto} 
but not always
made prominent\cite{hbyang,parker2010}. 
For instance, while the focus of ref.~\cite{hbyang}
was on particle-hole symmetry, they also presented evidence for asymmetry
which they analyzed further in ref.~\cite{yrzarpes} to find good agreement
with the ansatz of Yang, Rice and Zhang\cite{yrz}. {In fig.~S2 of
the supplemental information of ref.~\cite{parker2010}, particle-hole asymmetry
is seen clearly in the pseudogap state, whereas particle-hole
symmetry at low energies is seen in the superconducting state.}
There is still controversy, however, and hence obtaining signatures of electron
pockets and the pseudogap Dirac point would help to clarify this debate.

On the issue of electron pockets, we note that the signal shown in 
fig.~\ref{fig5}(d)-(f) is very large  and so STS should be able to confirm the
existence of these pockets. 
Given that they are associated with pockets in the antinodal regions of the
Brillouin zone, they should not be obscured by potential matrix element
effects, which are proposed to impact the nodal region.
STS experiments allow us to  
access another part of the band structure which has
been split by the pseudogap, and our results here indicate the
dominance of the electron pocket signal over that of the hole pocket
in this regime. This may be analogous to 
the quantum oscillations seen at high
magnetic field
in 
YBa$_2$Cu$_3$O$_{6.5}$ which 
 have been a cause for considerable debate due to the inferred presence of
electron pockets\cite{louis2} but a surprising
 lack of observation of those due to holes.

In summary, we have shown that signatures of electron pockets could be detected in FT-STS at
positive bias which would confirm the reconstruction of the Fermi surface
due to the formation of the pseudogap. Furthermore, 
AC-ARPES and FT-STS techniques 
can be used in a novel way to identify the pseudogap Dirac point which would
both confirm a two gap scenario for the high $T_c$ cuprates and allow for
detailed testing of candidate proposals for the pseudogap in the cuprates.
Some existing ARPES data 
presents suggestive evidence, although somewhat indirect, that a 
 pseudogap Dirac point is indeed present, at
reasonable values of energy and momentum,
 and motivates further investigation particularly involving QPI FT-STS
at positive bias.

\acknowledgments
We thank Johan Chang, J.C. Davis and J.P.F. LeBlanc for helpful discussions.
This work has been supported by the Natural Sciences and Engineering Council
of Canada (NSERC)  
and by the Canadian Institute for Advanced Research (CIFAR).

\end{document}